\documentclass{article}
\usepackage{microtype,graphicx,booktabs,array,amsmath,amssymb,xspace,float}
\usepackage{hyperref}
\usepackage[accepted]{mlsys2025}
\makeatletter
\renewcommand{\mlsys@appearing}{Code: \url{https://github.com/RegiaYoung/OOM-MoE}.}
\makeatother
\hypersetup{pdfsubject={Full-parameter mixture-of-experts training}}
\usepackage{refcount}
\newcommand{\sys}{OOM\xspace}
\newcolumntype{L}[1]{>{\raggedright\arraybackslash}p{#1}}

\providecommand{\Description}[1]{}
\mlsystitlerunning{OOM: Offload-Optimized Full-Parameter MoE Training via Memory-Execution Co-Design}

\makeatletter
\g@addto@macro\UrlBreaks{\do a\do b\do c\do d\do e\do f\do g\do h\do i\do j\do k\do l\do m\do n\do o\do p\do q\do r\do s\do t\do u\do v\do w\do x\do y\do z\do A\do B\do C\do D\do E\do F\do G\do H\do I\do J\do K\do L\do M\do N\do O\do P\do Q\do R\do S\do T\do U\do V\do W\do X\do Y\do Z\do 0\do 1\do 2\do 3\do 4\do 5\do 6\do 7\do 8\do 9}
\makeatother
\begin{document}
\twocolumn[
\mlsystitle{OOM: Offload-Optimized Full-Parameter MoE Training\texorpdfstring{\\}{ }via Memory-Execution Co-Design}
\begin{mlsysauthorlist}
\mlsysauthor{Ruijia Yang}{hkustgz}
\mlsysauthor{Shiyuan Lin}{hkustgz}
\mlsysauthor{Chengxi Liao}{hkustgz}
\mlsysauthor{Zeyi Wen}{hkustgz}
\par\smallskip
{\normalfont\small\ttfamily
\{ryang379,slin954,cliao118\}@connect.hkust-gz.edu.cn\\
wenzeyi@hkust-gz.edu.cn}
\end{mlsysauthorlist}
\mlsysaffiliation{hkustgz}{The Hong Kong University of Science and Technology (Guangzhou), Guangzhou, China}
\mlsyscorrespondingauthor{Zeyi Wen}{wenzeyi@hkust-gz.edu.cn}

\mlsyskeywords{Mixture of experts, full-parameter training, memory management}
\vskip 0.3in
\begin{abstract}
Offloading model state to host memory and streaming layer weights let a single multi-GPU node train large mixture-of-experts (MoE) models. GPU memory then becomes a performance budget, shared by resident expert state, retained activations, and execution worksets. The best allocation depends on the expert execution path, batch size, and sequence length. We present \sys (Offload-Optimized MoE), a full-parameter MoE training system that co-designs expert execution and memory allocation. For expert-parallel execution, \sys updates resident experts on their owning GPUs and jointly plans state residency and MoE activation retention according to pipeline dependencies. For local execution, \sys bounds GPU worksets through dependency-aware partitioning; in hybrid models, it also retains attention results on the host to avoid costly recomputation and selectively regenerates layer boundaries through cheaper layers to make room. Calibrated cost models guide joint configuration within GPU and host budgets to minimize step time; \sys selects among the configured paths through short executions. On four H100s, \sys trains with 1M-token contexts, whereas FSDP2, ZeRO-Offload, and MegaTrain exhaust GPU memory at 512K. It also trains a 122B-parameter MoE model, on which all three exhaust host memory. On four 24\,GB RTX 4090s, it trains 512K-token contexts and 1,024 1K-token sequences per step without gradient accumulation. Across matched-global-batch comparisons, \sys delivers $1.58\times$ and $4.08\times$ geometric-mean (up to $2.14\times$ and $7.77\times$) throughput over CPU-offloaded FSDP2 on the two platforms while using the least host memory among the evaluated systems---as little as $1/3.0$ of FSDP2's.
\end{abstract}

]
\printAffiliationsAndNotice{}
\section{Introduction}
\label{sec:intro}

\begin{figure}[t]
\centering
\includegraphics[width=\linewidth]{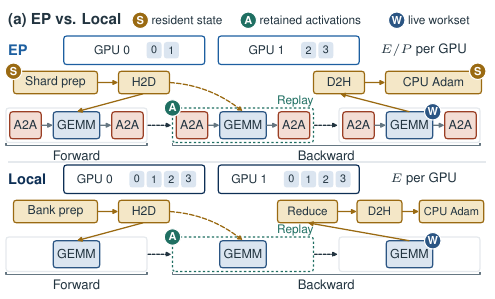}\\[0pt]
\includegraphics[width=\linewidth]{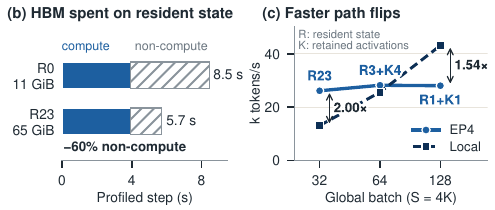}
\caption{Expert execution and GPU-memory value: (a) EP versus Local; (b) residency reduces step time; (c) configured-path crossover. (b,c) use four H100s at 4K context.}
\label{fig:intro-pipeline}
\end{figure}

Mixture-of-experts (MoE) models expand model capacity by activating only a few experts per token~\cite{fedus2022switch,jiang2024mixtral}. Qwen3.6-35B-A3B, for example, selects eight of 256 routed experts per token~\cite{qwen2026qwen36config}. Yet full-parameter training must maintain model state for all experts. This gap between sparse computation and total training state makes GPU memory (HBM) the bottleneck of full-parameter training.

State offloading and layer streaming keep model state in host memory and materialize working weights on GPUs as layers execute~\cite{ren2021zerooffload,rajbhandari2021zeroinfinity,yuan2026megatrain,yang2026slideformer,yang2026slidedp}. GPU memory then becomes a \emph{performance budget} with competing uses: retaining state to reduce host service, preserving activations to avoid recomputation, and supporting larger execution worksets. With four-way expert parallelism on four H100s at global batch 32 and 4K context, retaining selected expert state on GPUs instead of streaming all experts cuts Qwen3.6's step time from 8.48 to 5.02\,s ($1.69\times$) without changing model computation; Figure~\ref{fig:intro-pipeline}b splits a profiled step of each plan into compute and non-compute time. Host capacity is also shared by model state, saved layer inputs, and retained operator results.

Expert execution and memory allocation constrain each other. Execution determines data movement, live worksets, and the dependencies among computation, transfers, and updates; memory allocation determines which execution configurations fit and changes which work delays step completion. The two must therefore be designed together and judged by the time of the whole training step, not by isolated savings in transfers or recomputation.

ES-MoE adapts expert offloading and pinning, and MoP specializes execution by component, but neither jointly budgets GPU memory between selective state residency and activation retention~\cite{kim2024esmoe,nguyen2026mop}. ProTrain jointly plans state and activations through chunk- and block-level policies, while Mist co-optimizes parallelism and memory with an efficient tuner that assumes identical layers~\cite{yang2026protrain,zhu2025mist}. These policies leave two host-backed MoE problems open. First, choosing expert-shard residency, MoE activation retention, and token-routing chunk sizes together requires weighing host-state service against replay with token exchange. Second, within each GPU, operator partitioning, result retention, and boundary regeneration must respect operator-specific dependencies and both memory budgets, dividing host capacity between layer inputs that are cheap to regenerate and attention results that are expensive to recompute.

We present \sys (Offload-Optimized MoE), a single-node multi-GPU system that co-designs expert execution and memory allocation for full-parameter MoE training. It supports two complementary expert-execution paths: expert parallelism (EP) and local execution (Local) (Figure~\ref{fig:intro-pipeline}a,c). EP assigns disjoint expert shards to GPUs, so each GPU materializes fewer expert weights per layer but exchanges routed tokens~\cite{lepikhin2021gshard}. Local instead materializes all expert weights for the current layer on each GPU to process local tokens, avoiding that exchange.

\textbf{C1: EP execution with state--activation budgeting.} In EP, each expert shard is both an execution unit and an optimizer update unit. This ownership lets us retain and update selected shards on their GPUs, reducing host transfers and CPU optimizer work while streaming the rest through a bounded pipeline. Retaining MoE activations competes for this budget but eliminates corresponding replay and token exchange. Residency changes which work remains exposed when computation, transfers, and updates overlap, and thus the step-time benefit of activation retention. A dependency-based EP planner allocates GPU memory jointly to resident state, retained activations, and routing worksets.

\textbf{C2: Dependency-aware hybrid activation execution.} Local amortizes expert-bank materialization over more tokens, but larger token workloads expand activation and operator worksets. Recurrent mixing, full attention, and expert computation have different dependencies; we partition each along compatible axes and coordinate lifetimes to bound GPU worksets. Fitting the workload is not enough: in hybrid models that interleave full attention with cheaper recurrent or sliding-window layers, long-context attention results grow linearly in size but require quadratic computation to reproduce~\cite{qwen2026qwen36config,yang2025gateddelta,gemmateam2026gemma4}. We retain these results on the host and, when host capacity is tight, regenerate selected layer boundaries through cheaper layers to make room. A Local planner jointly chooses partitions, result retention, and boundary regeneration within GPU and host budgets. At 512K on four H100s, the combined design delivers $1.26\times$ the throughput of the partitioned baseline at nearly the same host footprint (Section~\ref{sec:eval:mechanisms}).

\textbf{C3: Workload-aware executable configuration.} Memory allocation can change EP and Local's relative performance, so we configure each eligible path before comparing step times. We jointly configure state placement and activation execution under a common step-time objective and memory budgets for each GPU and host NUMA node. Calibrated cost models guide within-path searches; where both paths are eligible, short executions select the plan with lower measured step time. Within one runtime, the best configured path is $1.1$--$2.0\times$ faster than the best unconfigured one. On four H100s at 4K context and global batch 96, EP4 is $1.17\times$ faster before configuration, yet Local is $1.21\times$ faster afterward (Section~\ref{sec:eval:selection}). Across 16 workloads on four models, short executions select the path that full-model runs measure as faster at every point.

We implement \sys in PyTorch and evaluate five MoE models with 8--256 experts and up to 122B parameters on H100 and RTX 4090 nodes. On four H100s, \sys trains Qwen3.6-35B-A3B with 1M-token contexts, the model's YaRN-extended length, whereas FSDP2, ZeRO-Offload~\cite{pytorch2026fsdp2,ren2021zerooffload}, and MegaTrain exhaust GPU memory at 512K. Beyond the four-GPU 1M-token workload, a single H100 completes a 2M-token training step with YaRN $\times$8. \sys also trains Qwen3.5-122B-A10B, whereas all three baselines exhaust the node's host memory. On four 24\,GB RTX 4090s, it completes 512K-token steps, a length at which the baselines exhaust 80\,GB H100s, and trains 1,024 1K-token sequences per step without accumulation; FSDP2 and ZeRO-Offload need it at that batch, and MegaTrain runs out of GPU memory.

Across matched-global-batch sweeps, \sys delivers $1.58\times$ and $4.08\times$ geometric-mean (up to $2.14\times$ and $7.77\times$) throughput over CPU-offloaded FSDP2 on H100 at 4K and RTX 4090 at 1K context ($2.92\times$ over ZeRO-Offload on RTX 4090) while using the least host memory among the evaluated systems---as little as $1/3.0$ of FSDP2's. Across four architectures at 1K context, it delivers $1.66$--$2.44\times$ FSDP2 throughput at batch 32 and $1.23$--$1.38\times$ at 256. On eight H100s at 1K context, it delivers $1.16$--$1.52\times$ the throughput of offloaded FSDP2 and $2.38$--$3.00\times$ that of Megatron-LM.

\section{Background and Motivation}
\label{sec:background}

\subsection{MoE Models and Training Memory}
\label{sec:background:state}
An MoE layer routes each token to $k$ of $E$ experts and combines their outputs~\cite{shazeer2017outrageously,jiang2024mixtral}. For equal-sized routed experts, the full bank contains $E/k$ times as many parameters as the experts selected for one token. Full-parameter training maintains state for every expert: FP32 master weights and two FP32 Adam moments alone require 12 bytes per parameter~\cite{kingma2014adam,ren2021zerooffload}, plus working weights and gradients. We call the masters and moments \emph{canonical state}; their updates produce the next working weights.

Activations and operator/communication worksets also consume memory, and their peak footprint depends on which objects are simultaneously live. The model's sequence-mixing operators also affect activation storage and recomputation costs: Mixtral and Qwen3 use attention throughout, Gemma4 interleaves sliding-window and global attention, and Qwen3.6 combines full attention with recurrent Gated DeltaNet (GDN) layers~\cite{jiang2024mixtral,qwen2025qwen3config,gemmateam2026gemma4,qwen2026qwen36card}.

\subsection{Training Beyond GPU Memory}
\label{sec:background:host}
ZeRO and FSDP shard state across GPUs; ZeRO-Offload and ZeRO-Infinity extend capacity through CPU and hierarchical memory~\cite{rajbhandari2020zero,zhao2023pytorchfsdp,ren2021zerooffload,rajbhandari2021zeroinfinity}. SlideFormer, SlideDP, and MegaTrain bound active parameter storage through layer streaming~\cite{yang2026slideformer,yang2026slidedp,yuan2026megatrain}. For streamed state, CPU masters supply working weights; backward uses the same weight version as forward, and aggregated gradients return for CPU updates. Transfers and updates overlap computation on other layers. Within this streaming framework, expert execution still changes weight delivery, token exchange, activation worksets, and replay costs. Bounding active parameter storage and overlapping host service therefore leave open which expert-execution path to use and how to divide GPU memory among resident state, retained activations, and worksets to reduce step time.

ES-MoE pipelines expert offloading and computation with token-load balancing; MoP combines component-specific parallelism with heterogeneous updates~\cite{kim2024esmoe,nguyen2026mop}. LazyTrain jointly schedules checkpoints, recomputation, and transfers~\cite{wu2026lazytrain}. KTransformers-FineTune (KT-FT) takes a different route, computing routed experts on the host CPU and replacing weight offload with CPU--GPU co-computation~\cite{li2026ktft}. ProTrain jointly tunes residency, buffers, checkpointing, and offloading, with GPU updates for persistent state; Mist co-optimizes memory and parallelism using overlap-aware modeling~\cite{yang2026protrain,zhu2025mist}. ProTrain's persistent chunks remain unpartitioned and Mist's efficient tuner assumes identical layers; neither policy space jointly configures expert-owned residency, MoE activation retention, and hybrid result retention.

\subsection{Execution Costs in Host-Backed MoE Training}
\label{sec:background:execution}
\label{sec:background:parallelism}
\label{sec:background:workload}
\label{sec:background:lifetimes}
EP partitions experts across $P$ GPUs, assigning $E/P$ per GPU under balanced placement, and exchanges tokens and results through all-to-all (A2A) communication~\cite{lepikhin2021gshard}. Local execution materializes the working weights of all $E$ experts in the current layer on each GPU to process local tokens, avoiding expert-token exchange while still requiring weight delivery and gradient aggregation. Optimizer state need not follow the placement of these working weights~\cite{skiadopoulos2026symi}. Retaining MoE activations in EP avoids the corresponding backward replay and its A2A; the retained activations compete with resident state and execution worksets for GPU memory.

At fixed sequence length, larger local batches amortize expert materialization over more tokens while increasing activation storage. Sequence length also changes the balance of operator arithmetic. For $B$ local sequences of length $S$, fixed architecture and routing degree give leading terms
\begin{equation}
F(B,S) \approx aBS+bBS^2,
\label{eq:workload-arithmetic}
\end{equation}
where projections, experts, and recurrent mixing contribute to $a$, and full attention to $b$. At equal token count $T=BS$, the full-attention term $bTS$ still depends on $S$, so batch size and sequence length must be modeled separately.

\begin{table}[t]
\caption{Leading arithmetic costs and operator dependencies at fixed widths and routing degree. $B$ counts local sequences and $S$ their length.}
\label{tab:operator-dependencies}
\centering
\small
\begin{tabular}{@{}L{.23\linewidth}L{.17\linewidth}L{.49\linewidth}@{}}
\toprule
Operator & Compute & Dependency \\
\midrule
GDN mixing & $O(BS)$ & Recurrent state across sequence segments \\
Full attention & $O(BS^2)$ & Causal token interactions within each head \\
Expert FFN & $O(BS)$ & Token-wise evaluation after routing \\
\bottomrule
\end{tabular}
\end{table}

Within each GPU, partitioning each operator along the axes its dependencies permit (Table~\ref{tab:operator-dependencies}) bounds worksets: GDN sequence chunks pass recurrent state~\cite{yang2025gateddelta}, full attention admits head groups over complete sequences, and expert FFNs admit token chunks after routing. Each partition must preserve backward state and gradient flow.

Checkpointing trades storage for recomputation~\cite{chen2016training,jain2020checkmate}; FlashAttention avoids quadratic score storage, and MoEBlaze reduces routing buffers with selective recomputation~\cite{dao2022flashattention,zhang2026moeblaze}. For dense LLMs, OOMB reaches million-token contexts by recomputing activations in chunks and offloading the KV cache~\cite{li2026oomb}. Long sequences expose an asymmetry: full attention's computation grows quadratically with length, while its output and softmax statistics grow only linearly, so keeping them can beat recomputation. Host memory bounds retention: decoder inputs (layer boundaries) and retained results compete with model state.

\section{System Overview}
\label{sec:overview}
\label{sec:overview-map}
\begin{figure*}[t]
\centering
\includegraphics[width=\textwidth]{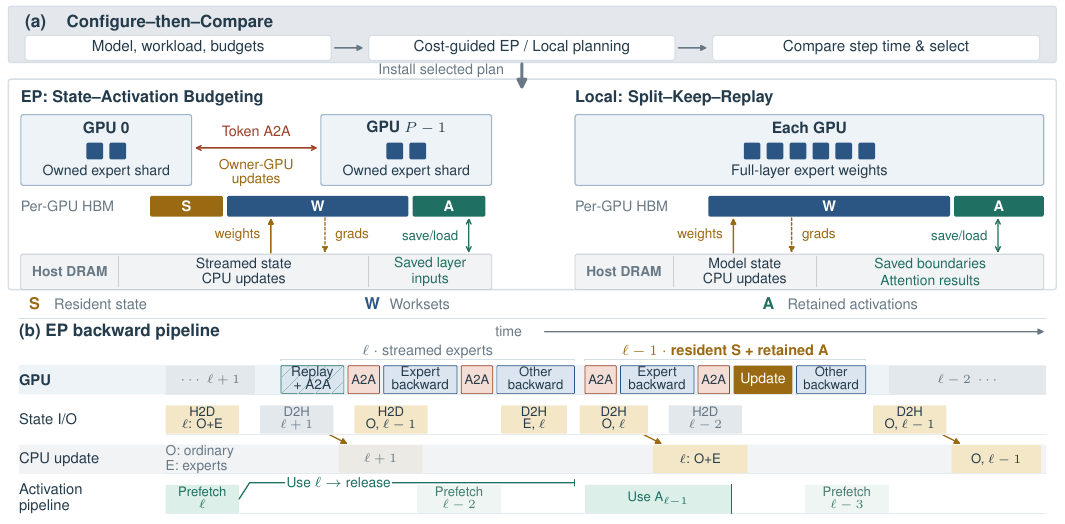}
\caption{Overview of \sys. (a) Cost-guided planning and EP/Local resource placement. EP and Local show alternative placements on the same node. (b) EP backward pipeline; O/E denote ordinary/expert components. Widths are schematic.}
\label{fig:system-overview}
\end{figure*}

\sys converts a workload and its memory budgets into a plan for a shared training runtime (Figure~\ref{fig:system-overview}). Host memory supplies capacity for model state and saved activations; bounded parameter windows feed GPU computation. Routed experts use either disjoint EP shards with token A2A or a complete bank serving local tokens. Attention, routers, and shared experts use the common parameter stream.

The backward pipeline overlaps state delivery, GPU computation, activation transfers, and host updates. Completion events preserve weight versions and release storage after its last consumer. GPU memory serves three purposes: resident state $M_{\mathrm{state}}$, retained activations $M_{\mathrm{keep}}$, and the active workset $M_{\mathrm{work}}$ (labeled $S$, $A$, and $W$ in the figure).

EP plans state residency and activation retention (Section~\ref{sec:ep-execution}); Local plans partitioning, result retention, and boundary regeneration (Section~\ref{sec:activation-execution}). Under a common step-time objective and GPU/NUMA budgets, Configure-then-Compare uses calibrated costs to configure each path and short executions to compare them (Section~\ref{sec:workload-execution}). Our PyTorch runtime~\cite{paszke2019pytorch} installs the plan before training.

\section{Expert-Parallel Execution and Memory Budgeting}
\label{sec:ep-execution}
An EP shard is both an execution unit and an update unit. \sys uses this ownership to keep selected experts' master weights, optimizer moments, and updates on their GPUs. The same memory can instead retain activations and remove replay, including its token routing. We budget memory by the work these choices remove from the training step.

\subsection{Expert Ownership and the Shared Pipeline}
Within an EP group, each GPU owns disjoint routed experts. A streamed shard keeps its FP32 master and Adam moments on the host: a worker prepares BF16 weights, and backward returns gradients for CPU update. Promoting a shard replaces host preparation, parameter H2D, gradient D2H, and CPU update with local weight generation and GPU AdamW. Each shard still has a single updater; resident and streamed experts both participate in token dispatch/return A2A. Attention, normalization, routers, and shared experts continue through the common stream.

\label{sec:ep-pipeline}
Figure~\ref{fig:system-overview} shows how this service overlaps computation within bounded parameter windows and gradient slots. Three dependencies preserve the execution: computation waits for its required weight version, including the original version during replay; updates wait for all required gradients; storage reuse waits for the last consumer. The GPU can therefore execute layer $\ell-1$ while the CPU updates layer $\ell$. Interleaving resident layers gives host updates additional GPU work to overlap, making both the number and positions of resident layers useful planning choices.

\subsection{Spend Memory on the Completion Path}
\label{sec:ep-budget}
\label{sec:pipeline-transform}
\label{sec:pipeline-dependencies}
Activation checkpointing supplies the complementary choice~\cite{chen2016training,korthikanti2023reducing}. Full-decoder checkpointing offloads the input and reconstructs internal tensors during backward. Splitting at the attention/MoE boundary allows either sublayer's backward tensors to remain on GPU; retaining both removes decoder-level replay. In EP, MoE retention saves expert recomputation \emph{and} the token exchanges required for replay.

The value of retention depends on what delays completion. Let $T_h$ be the duration of a submitted host service and $T_g$ the independent GPU work before its consumer. If that service is the final prerequisite, the exposed wait is $[T_h-T_g]_+$. Residency shortens host service and thereby changes which work is exposed, so retention must be valued against the resulting critical path. Joint budgeting scores complete candidates rather than adding independent gains.

For candidate $z$, the per-GPU peak is
\begin{equation}
\begin{split}
M^{\mathrm{peak}}_{\mathrm{GPU}}(z)=M_0+M_{\mathrm{state}}(z)\\
+\max_t\!\left\{M_{\mathrm{keep}}(t,z)+M_{\mathrm{work}}(t,z)\right\}+\epsilon,
\end{split}
\label{eq:ep-live-budget}
\end{equation}
where $M_0$ is fixed occupancy and $\epsilon$ is headroom. The workset includes working weights, gradients, routing buffers, and operator temporaries. CPU checkpoint inputs consume host capacity. State cost depends on expert parameters and EP partitioning, whereas activation and routing costs grow with token load. Smaller routing chunks free memory for residency or retention but require more communication rounds. Consequently, batch size and sequence length determine how much memory each purpose deserves.

\subsection{Produce an Executable EP Plan}
\label{sec:ep-planner}
\label{sec:pipeline-policy}
\label{sec:pipeline-returns}
For a fixed EP layout, the planner selects resident layers $R$, checkpoint policies, and routing chunk $c$. It explores contiguous and interleaved resident sets; the calibrated checkpoint family retains MoE backward tensors in a set $K$ and fully checkpoints other layers. Reference occupancy and incremental costs filter candidates using Equation~\ref{eq:ep-live-budget}.

The time model combines calibrated forward work, a dependency-simulated backward region, and the final update drain. For a stage $v$ with service time $d_v$ and predecessor set $\mathcal P_v$, completion follows
\begin{equation}
f_v=d_v+\max_{u\in\mathcal P_v}f_u.
\label{eq:ep-stage-completion}
\end{equation}
Edges encode data readiness, stream order, and buffer reuse. Thus replay, routing, transfers, and updates overlap only where execution permits; waiting emerges from dependencies instead of being counted again as service. The fastest predicted feasible candidate determines the plan.

As activation demand grows with batch, the configured EP4 plan shifts memory from state to activations: 23 resident layers at global batch 32, three resident and four MoE-retained layers at 64, and one of each at 128 (Figure~\ref{fig:intro-pipeline}c).

\section{Local Execution with Split--Keep--Replay}
\label{sec:activation-execution}
Local execution avoids token exchange and amortizes a full expert bank over local tokens. Its advantage grows with token load, making activation capacity central. \sys manages hybrid activations with \emph{Split--Keep--Replay}: split each operator along axes its dependencies permit, keep expensive results, and regenerate selected boundaries by replaying cheaper decoders. A Local planner coordinates these choices within GPU and host budgets.

\subsection{Deliver the Expert Bank}
\label{sec:local-state}
As in layer-streaming systems~\cite{rajbhandari2021zeroinfinity,yuan2026megatrain,yang2026slideformer,yang2026slidedp}, Local keeps canonical state on the host and materializes each layer's full expert bank in a double-buffered parameter window. \emph{Full} delivery copies a shared BF16 bank to each rank; \emph{Shard} delivers disjoint slices and reconstructs the bank through GPU AllGather, cutting each rank's parameter H2D to about $1/P$ for $P$ GPUs. Gradients are reduced on the GPUs or in host shared memory, accumulating BF16 partials in FP32 before the CPU update.

\subsection{Partition along Dependency-Compatible Axes}
\begin{figure}[t]
\centering
\includegraphics[width=\columnwidth]{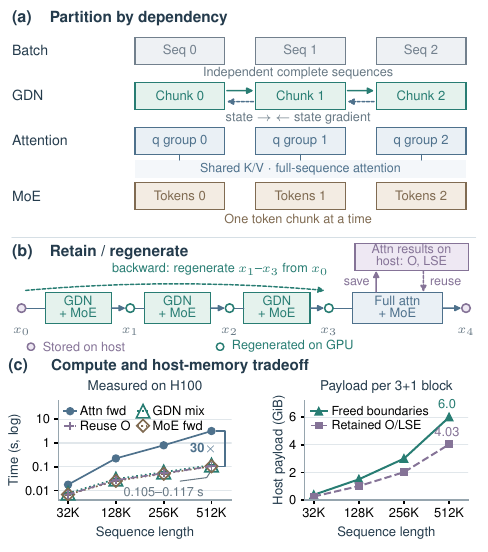}
\caption{Hybrid activation execution: (a) dependency-aware partitioning; (b) retention and regeneration; (c) Qwen3.6 operator times on H100 and calculated host payloads per sequence.}
\label{fig:hybrid-execution}
\end{figure}

Large batches admit independent sub-batches of complete sequences. Long sequences require different partitions for recurrent mixing, attention, and expert computation (Figure~\ref{fig:hybrid-execution}(a)). Each partition bounds temporary storage, preserves forward/backward dependencies, and accumulates parameter-gradient partials in FP32.

\paragraph{GDN: sequence chunks with state handoff.}
Chunks pass an FP32 recurrent state and causal-convolution left context~\cite{yang2025gateddelta,fla2024}. Backward rematerializes chunks in reverse, passing the final-state derivative into each chunk and propagating its initial-state derivative to the preceding one. Convolution-context gradients join the preceding slice. For chunk length $C$, $H_v$ value heads, and $d_k\times d_v$ state matrices, one GDN layer's chunk-boundary states occupy
\begin{equation}
M_{\mathrm{GDN\ states}}=4B(\lceil S/C\rceil-1)H_vd_kd_v\quad\text{bytes}.
\label{eq:gdn-state}
\end{equation}
Smaller chunks reduce token workspace but increase state storage and calls. Complete-sequence sub-batches provide an additional control for large $B$.

\paragraph{Full attention: head groups over complete sequences.}
\sys splits the query heads of each GQA group, which share one K/V head, into groups of $g$. Each group retains the full causal sequence, materializes its projections and intermediates, and releases them after use. Backward rebuilds groups and accumulates shared K/V and input gradients in FP32. For head dimension $d$ and sub-batch size $B_{\mathrm{sub}}$, projection and gate workspace scales with $B_{\mathrm{sub}}Sgd$. Head grouping complements FlashAttention's score-matrix tiling~\cite{dao2023flashattention2} by bounding projection workspace.

\paragraph{MoE: token chunks.}
Normalization, routing, routed experts, and the shared branch process token slices; backward reconstructs and releases one slice at a time. Chunking bounds temporaries, while input, output, and gradient buffers still scale with $BS$.

\subsection{Retain Expensive Results, Regenerate Boundaries}
\label{sec:activation-lifecycle}
Hybrid models make recomputation costs sharply asymmetric: attention arithmetic grows quadratically with sequence length, while its output grows only linearly. At 512K, attention forward takes 3.180\,s, versus 0.117\,s for GDN mixing and 0.105\,s for MoE, and costs $30\times$ its replay with output reuse, up from $2.6\times$ at 32K (Figure~\ref{fig:hybrid-execution}(c), left).

\sys therefore saves attention output $O$ and softmax log-sum-exp statistics (LSE) on the host. Checkpoint replay reuses $O$, reducing attention-sublayer replay to 0.105\,s; backward consumes $O$/LSE while rebuilding Q/K/V. Reuse across both phases removes redundant attention forwards from nested rematerialization inside layer checkpointing~\cite{chen2016training,jain2020checkmate,korthikanti2023reducing}. Selective recomputation discards large yet cheap-to-rebuild attention activations at moderate lengths~\cite{korthikanti2023reducing}. Long context reverses that trade: \sys retains attention's linear-size results and rebuilds only cheap projections during replay.

Decoder inputs compete for the same host capacity: $L$ BF16 boundaries of hidden width $h$ occupy $2LBSh$ bytes. \sys keeps selected boundaries and regenerates the others with no-gradient forwards from the nearest kept input. In Figure~\ref{fig:hybrid-execution}(b), keeping $x_0$ and $x_4$ around three GDN decoders and a full-attention layer recovers $x_1$--$x_3$ through the three complete GDN decoders, including normalization and MoE, without repeating attention. At 512K, these three boundaries free 6.0\,GiB of host memory per sequence, more than the 4.03\,GiB of $O$/LSE (Figure~\ref{fig:hybrid-execution}(c), right). Sliding-window hybrids show the same asymmetry: Gemma4-26B-A4B interleaves five 1K-window layers with each global-attention layer; at 256K, replay with $O$ reused cuts a global-attention forward from 17.2 to 0.04\,s, and one block's five boundaries free 6.9\,GiB per sequence against 4.0\,GiB of $O$/LSE.

Regenerated boundaries stay on GPU for reverse-order consumption; a cap on their number bounds the footprint, with spare GPU memory permitting longer runs. The planner weighs saved attention replay and freed host capacity against whole-decoder regeneration, temporary GPU storage, and changes in exposed transfer time. Ample host capacity allows both boundaries and results to remain stored; tighter budgets favor regenerating cheap boundaries to retain expensive attention results. The regeneration interval is a planned quantity set by sequence length and both budgets.

\subsection{Coordinate Lifetimes and Plan the Workset}
Operator workspaces coexist with recurrent states, regenerated boundaries, and prefetched inputs. \sys schedules them around consumption. The original GDN forward rolls its state; backward boundary states are created during the current layer's replay and released after use. Activation-prefetch buffers are allocated on demand. Grouped attention defers input prefetch until query-head workspace is released, then overlaps transfer with remaining backward computation. Regeneration uses original-version weights from the shared parameter window. Together, these choices turn local partitioning~\cite{zhao2026memfine,pandit2026flattening} into control over the step's simultaneous peak.

\label{sec:hybrid-planner}
The Local planner searches these coupled choices jointly: partitions set each phase's workspace, which competes with held boundaries for GPU memory, while $O$/LSE competes with kept boundaries for host memory. It enumerates attention head groups, GDN sequence/sub-batch chunks, and MoE token chunks. For each combination, a layerwise dynamic program chooses boundary policies and attention-result placement. Retaining a boundary charges its storage; dropping it extends a regeneration run and charges the preceding decoder's forward. Retaining attention results replaces replay work with storage and transfer costs.

DP states preserve open regeneration runs, GPU objects carried across layers, and pending transfer dependencies. Labels record accumulated time, aggregate host storage, and the running GPU peak. Within the same DP state, a label is discarded if another is no slower and uses no more of any resource. GPU peak is the largest live footprint across execution phases, including held boundaries and the output head, rather than the sum of separate workspace maxima. Surviving plans are checked against each NUMA node's host budget, including model state and pinned pools, and the fastest predicted feasible plan is selected.

The resulting plan specifies partitions, retained boundaries, result placement, and regeneration limits; FlashAttention, FLA, and Liger adapters execute it~\cite{dao2023flashattention2,fla2024,hsu2024ligerkernelefficienttriton}.

\section{Configure-then-Compare}
\label{sec:pipeline-planning}
\label{sec:workload-execution}
\begin{figure}[t]
\centering
\includegraphics[width=\columnwidth]{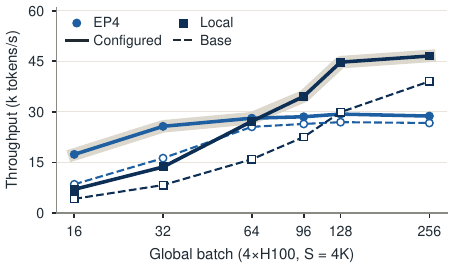}
\caption{EP4/Local throughput before and after configuration. Gray traces the upper measured Configured curve.}
\label{fig:envelope}
\end{figure}

EP and Local convert memory into throughput differently, so \sys applies \emph{Configure-then-Compare}: it first allocates resources within each path, then selects among them. Path choice thus reflects the execution each memory budget enables. Configuration raises each path's throughput and can change which path is faster: the upper envelope in Figure~\ref{fig:envelope} marks the faster configured path at each batch size.

\subsection{Model the Complete Configuration}
The input supplies model structure, local batch $B$, sequence length $S$, precision, GPU topology, and resource budgets. A configuration $z$ specifies expert execution and state delivery, resident layers, checkpoint and boundary policies, operator partitions, and result placement. For the supported candidates $\mathcal C(w)$ at workload $w$, the objective is
\begin{equation}
\begin{aligned}
z^*={}&\arg\min_{z\in\mathcal C(w)}T_{\mathrm{step}}(z),\\
\text{s.t.}\quad&M^{\mathrm{peak}}_{\mathrm{GPU},i}(z)\le G_i,\\
&M^{\mathrm{peak}}_{\mathrm{host},n}(z)\le H_n,
\end{aligned}
\label{eq:configuration-target}
\end{equation}
where $G_i$ and $H_n$ are the memory budgets of GPU $i$ and NUMA node $n$. Time includes forward, replay, backward, gradient processing, optimizer updates, and weight publication. Memory follows simultaneous allocations, including transfer staging. This step-level accounting therefore values residency, retention, and partitioning consistently.

\paragraph{Calibrate service and storage costs.}
Short reference runs on a few repeating decoder units characterize each path's compute, replay, communication, and host services, as well as reference occupancy. Matching component profiles can be reused. EP composes these costs through execution dependencies (Section~\ref{sec:ep-planner}); Local sums operator costs, regeneration, and exposed transfers, and checks memory against its phase-wise live set (Section~\ref{sec:hybrid-planner}). Keeping $B$ and $S$ separate captures both token-scaled work and the sequence-dependent cost of attention.

\begin{figure*}[t]
\centering
\includegraphics[width=\textwidth]{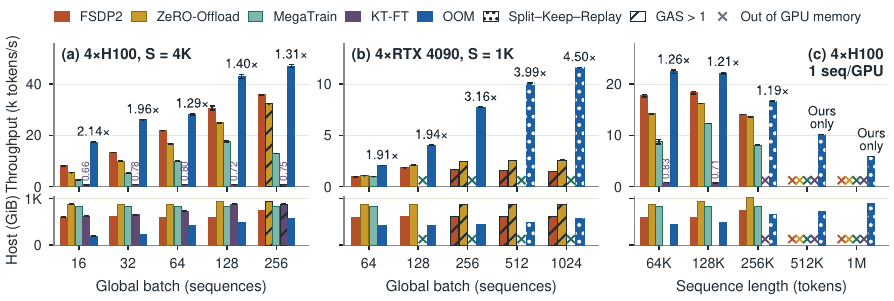}
\caption{Batch-size and context-length scaling. Labels give speedup over the fastest baseline; labels on KT-FT bars give k tokens/s.}
\label{fig:release-e2e}
\label{fig:long-baselines}
\label{fig:batch-sequence}
\end{figure*}

\paragraph{Search within each path.}
State-delivery mode, transfer chunk sizes, and buffer settings are supplied runtime settings; their service and storage costs enter calibration. Within these settings, EP planning chooses residency, checkpointing, and routing chunks by its dependency model or, where none is calibrated, from measured per-layer costs; Local enumerates operator partitions and searches boundary/result placement. Each planner filters capacity violations, ranks the remaining candidates by predicted step time $\widehat T_{\mathrm{step}}$, and emits the settings required to execute its preferred plan on the shared runtime.

\subsection{Compare Configured Paths}
\label{sec:workload:select}
For $N=BS$ tokens per GPU, top-$k$ routing, width $h$, and $b$-byte elements, balanced EP routing transmits
\begin{equation}
V_{\mathrm{EP}}=2kNhb\frac{P-1}{P}
\label{eq:ep-token-volume}
\end{equation}
bytes per GPU for dispatch and return in one MoE pass. Local's expert-bank delivery is independent of $N$, so more tokens amortize its materialization cost. Calibrated models price these transfers with overlap, replay, and live worksets.

Long-context workloads use Local's Split--Keep--Replay planner, built for their activation worksets. Where both paths are candidates, each planner proposes its best feasible plan, and short executions at the same workload and budgets select the faster one: prediction narrows the space, and execution resolves the cross-path comparison. Because decoder layers repeat, each path first runs one and two repeating units, which separates its fixed cost from its per-layer cost and reads its GPU occupancy; EP then runs its plan's resident and retained shares once more. Scaling these tests to the full depth predicts each configured path's step time at the target workload. The runtime then installs the chosen path, its parameter windows, retained objects, and completion dependencies before training begins.

\section{Evaluation}
\label{sec:eval}
We measure throughput, host memory, and capacity, then isolate the mechanisms and configuration choices behind them and test training quality and model coverage.

\subsection{Experimental Setup}
We train the 40-layer Qwen3.6-35B-A3B language backbone (256 top-8 experts, 30 GDN and 10 full-attention layers) with AdamW on packed OpenThoughts3 inputs~\cite{guha2026openthoughts}, updating all language-model parameters. We also evaluate Mixtral-8x7B, Qwen3-30B-A3B, Gemma4-26B-A4B, and Qwen3.5-122B-A10B. We write R$n$ and K$n$ for $n$ resident expert layers and $n$ MoE-retained layers.

\paragraph{Baselines and resources.}
Our baselines span sharded offload (FSDP2 and DeepSpeed ZeRO-3 with CPU parameter/optimizer offload, \emph{ZeRO-Offload}), layer streaming (MegaTrain), CPU co-computation (KT-FT), and model parallelism with CPU optimizer offload (Megatron-LM)~\cite{pytorch2026fsdp2,ren2021zerooffload,yuan2026megatrain,li2026ktft,shoeybi2019megatron,nvidia2026bridge}. FSDP2 and \sys share FA3 attention~\cite{shah2024flashattention3} and Liger MoE kernels in the main four-H100 sweeps, retaining their own memory policies. All use BF16 compute and FP32 master weights and moments. FSDP2 and Megatron reduce gradients in FP32; ZeRO-Offload communicates BF16 gradients. KT-FT runs its native AMX BF16 CPU experts on the H100 host; we add the standard FP32 masters and moments of the other systems in place of its native BF16 optimizer state. We found no public implementation of ProTrain, and released Mist and ES-MoE code lacks native support for the recent MoE models evaluated here.

The H100 server has eight 80\,GB GPUs on NVLink and two Xeon 8468 sockets with 1,008\,GiB each; the RTX 4090 server has four 24\,GB GPUs on PCIe and two Xeon 6338N sockets with 503\,GiB each. Each comparison runs all systems on the same GPUs and host. \sys uses no gradient accumulation. Baselines accumulate when needed to match the global batch, keeping long sequences intact.

\paragraph{Measurement.}
Main sweeps use two independent processes, each with two warmup and five timed training steps. We average their medians of slowest-rank step times; whiskers span the two medians. Split--Keep--Replay studies use one process with one warmup and three measured steps. Step time covers computation, replay, transfers, reductions, optimizer updates, and weight publication. Throughput counts input tokens once; sweep summaries use geometric means of paired ratios. Resource runs sample process-tree PSS and peak GPU allocation; 512K/1M and Split--Keep--Replay host footprints sum rank-level PSS estimates.

\subsection{Throughput Across Batch Sizes}
\label{sec:eval:e2e}
On four H100s at 4K context, \sys delivers $1.29$--$2.14\times$ FSDP2 throughput across global batches 16--256 ($1.58\times$ geometric mean; Figure~\ref{fig:release-e2e}a), $1.45$--$3.16\times$ that of ZeRO-Offload, and $2.46$--$6.63\times$ that of MegaTrain, reaching 47.0K tokens/s. EP4 serves batches 16--64 (R27, R23, and R3/K4) with smaller banks and resident state; Local serves 128--256, amortizing expert banks without token A2A. At batch 32, \sys runs $4.8\times$ faster than Megatron-LM, which needs eight accumulation rounds. KT-FT reaches 659--803 tokens/s across batches 16--128, so \sys is $26$--$60\times$ faster; at batch 256 KT-FT needs two accumulation rounds (751 tokens/s, $63\times$). Routed experts are dense matrix computation: even on a 48-core AMX Xeon, KT-FT's CPU expert path takes 70\% of its exposed step time at batch 32, against 4\% for the FP32 optimizer update.

On 24\,GB RTX 4090s, the gains grow to $2.16$--$7.77\times$ over FSDP2 and $1.91$--$4.50\times$ over ZeRO-Offload across batches 64--1024, with geometric means of $4.08\times$ and $2.92\times$ (Figure~\ref{fig:release-e2e}b). At batch 1024, each GPU processes 256 complete 1K sequences without accumulation (11.6K tokens/s), whereas both baselines need eight accumulation rounds. Split--Keep--Replay's partitioning enables this amortization: default operators fit only 64 sequences per GPU, and the Local planner shifts GDN/MoE chunks from 32K/16K tokens at 128 sequences per GPU to 16K/32K at 256, where keeping 32K/16K exhausts GPU memory.

\paragraph{Memory as a performance budget.}
Residency spends GPU capacity to reduce host service and storage; Local leaves more GPU space for the workset. Across H100 batches, \sys uses 197--571\,GiB of host memory versus FSDP2's 591--751\,GiB and KT-FT's 619--876\,GiB, with resident experts reducing batches 16/32 to 197/230\,GiB (Figure~\ref{fig:release-e2e}a). On RTX 4090, \sys uses 0.70--0.93$\times$ FSDP2's host memory and 0.48--0.66$\times$ ZeRO-Offload's; at 64K--256K on H100, it uses 0.76--0.87$\times$ FSDP2's (Figure~\ref{fig:release-e2e}b,c).

\subsection{Long-Sequence Efficiency and Capacity}
\label{sec:eval:scaling}
With one sequence per GPU, \sys trains at 22.4K, 22.2K, and 16.7K tokens/s for 64K, 128K, and 256K contexts, $1.19$--$1.26\times$ the throughput of FSDP2 (Figure~\ref{fig:release-e2e}c), $1.22$--$1.58\times$ that of ZeRO-Offload, and $1.79$--$2.57\times$ that of MegaTrain. KT-FT reaches 834 and 714 tokens/s at 64K and 128K, $27$--$31\times$ slower, and exhausts GPU memory at 256K. Our default operators suffice up to 128K; from 256K, Split--Keep--Replay is faster and saves GPU memory.

Split--Keep--Replay extends this range to 512K and 1M: with grouped attention, operator chunks, $O$/LSE retention, and boundary regeneration, \sys completes a step in 205.7 and 712.6\,s (10.2K and 5.9K tokens/s at 733 and 894\,GiB of host memory; YaRN~\cite{peng2024yarn} $\times$2 and $\times$4), whereas FSDP2, ZeRO-Offload, and MegaTrain exhaust GPU memory at both lengths. One H100 also completes 2M-token steps at 805 tokens/s (YaRN $\times$8).

\subsection{State and Activation Mechanisms}
\label{sec:eval:mechanisms}
\label{sec:eval:joint}
\paragraph{Resident experts remove host service.}
At batch 32 and 4K context, the planner's R23 plan with a 64K routing chunk cuts step time from 8.48 to 5.02\,s ($1.69\times$; predicted 5.14\,s) without manual tuning, raising GPU allocation from 11.4 to 64.8\,GiB for resident state. Profiled steps (8.5 and 5.7\,s; Figure~\ref{fig:intro-pipeline}b) show why: computation with replay stays at 3.93\,s, while non-compute time falls by 60\%, from 4.55 to 1.80\,s, as expert-owned updates remove host preparation, transfers, and CPU optimization.

\paragraph{Placement and capacity shape the gain.}
Position also shortens the critical path. With the same 12 resident layers at batch 32 and 4K, interleaving them cuts step time from 7.53 to 6.69\,s ($1.13\times$; predicted $1.14\times$) at the same GPU and host memory. The two plans do the same GPU and CPU work and differ in exposed wait (Section~\ref{sec:ep-budget}): consecutive resident layers run after the streamed layers' updates have already stalled the GPU, whereas interleaved ones give those updates GPU work to overlap. At batch 32 and 1K, step time falls from 8.11\,s without resident layers to 3.54\,s at the planned R27 ($2.29\times$), and the gain lasts until capacity runs out: on Gemma4 at batch 128, the planned R18 is within 0.3\% of the fastest point tested, whereas R22 leaves no GPU headroom and runs $1.35\times$ slower.

\paragraph{Retention's value depends on residency.}
Retaining one decoder layer's MoE activations removes its MoE replay and token exchange, shortening the step by 5.2\% with 17 resident layers but by under 1\% with 21. Residency changes the exposed work and retention's benefit, so the EP planner budgets the two jointly (Section~\ref{sec:ep-planner}). At about 64\,GiB, its R23/K0 plan (5.02\,s) beats R21/K1 (5.33\,s).

\begin{figure}[t]
\centering
\includegraphics[width=\columnwidth]{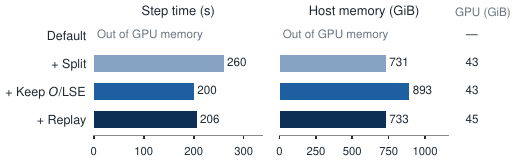}\\[0pt]
{\small (a) Cumulative mechanisms at 512K, four H100s}\\[1pt]
{\small
\setlength{\tabcolsep}{5pt}
\begin{tabular}{@{}lrrrr@{}}
\toprule
 & \multicolumn{2}{c}{Context, 1 seq/GPU} & \multicolumn{2}{c}{Batch per GPU} \\
\cmidrule(lr){2-3}\cmidrule(l){4-5}
GPU & Default & S--K--R & Default & S--K--R \\
\midrule
H100 & 256K & 1M & 64 & 256 \\
RTX 4090 & 64K & 512K & 64 & 256 \\
\bottomrule
\end{tabular}\\[0pt]
(b) Largest completed workload, four GPUs}
\caption{Split--Keep--Replay (S--K--R) gains and capacity. Batch contexts: 4K on H100; 1K on RTX 4090.}
\label{fig:c2-ablation}
\end{figure}

\paragraph{Partition, retain, and regenerate.}
Figure~\ref{fig:c2-ablation} shows how Split--Keep--Replay turns GPU and host capacity into throughput. At 512K with one sequence per GPU, the default path exhausts GPU memory in its first MoE backward. Dependency-compatible partitioning and lifetime coordination make the step fit in 43.4\,GiB at 260.0\,s. Retaining attention $O$/LSE on the host avoids recomputing the attention core and cuts step time to 200.5\,s, a $1.30\times$ gain, at the cost of 161\,GiB more host memory.

Regenerating half of the decoder boundaries recovers 160\,GiB of it for 2.8\% more step time and 2\,GiB of GPU memory. The complete design thus runs $1.26\times$ faster than the partitioned baseline at nearly the same host footprint (733 versus 731\,GiB). At 256K, where the default path fits, partitioning with $O$/LSE retention still runs $1.06\times$ faster (62.8 versus 66.9\,s) and lowers peak GPU allocation by 27\%, from 53.8 to 39.4\,GiB. Split--Keep--Replay thus improves feasible workloads as well as extending capacity: on both platforms it raises the largest completed context by 4--8$\times$ and the per-GPU batch by 4$\times$ (Figure~\ref{fig:c2-ablation}b). Default execution exhausts GPU memory at the next doubling.

\begin{figure}[t]
\centering
\includegraphics[width=\columnwidth]{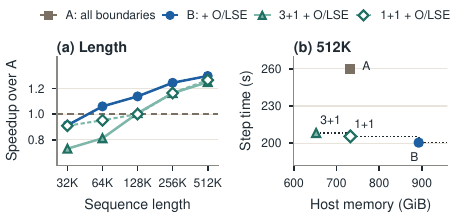}
\caption{Split--Keep--Replay policies on four H100s. Dotted line: fastest measured policy within each host-memory budget.}
\label{fig:c2-policies}
\end{figure}

\paragraph{The best policy depends on length and budget.}
Policy A keeps every decoder boundary on the host and recomputes attention; B also keeps $O$/LSE; 1+1 and 3+1 add regeneration, keeping one boundary in every two and four layers. Within partitioned execution, retaining $O$/LSE adds 9\% to step time at 32K but delivers $1.14\times$ throughput at 128K and $1.30\times$ at 512K (Figure~\ref{fig:c2-policies}a), as attention computation outgrows its retained results; from 128K, relative regeneration overhead likewise shrinks with length. At 512K, B is fastest at 893\,GiB; 1+1 saves 160\,GiB for under 3\% more time, and 3+1 saves 240\,GiB for 3.8\% (Figure~\ref{fig:c2-policies}b). Compared with A, 3+1 is both $1.25\times$ faster and 79\,GiB smaller: inexpensive boundaries make room for valuable attention results. Predictions frozen before execution rank these policies in measured order, within 0.6\% of step time and 0.2\,GiB of host memory, so at every host budget the predicted-fastest feasible policy is the measured-fastest one. Under a 750\,GiB budget, the planner holds the last 17 boundaries on GPU and runs in 202.1\,s, faster than 1+1 (205.5\,s), the fastest fixed policy that fits this budget.

\subsection{Workload-Aware Configuration}
\label{sec:eval:selection}
\label{sec:eval:analysis}
Configured EP4 shifts memory from state to activations as the batch grows (Section~\ref{sec:ep-planner}), yet at batch 128 Local is $1.54\times$ faster (12.18 versus 18.71\,s), a gain fixed EP forgoes. Across batches 16--256, the best configured path is $1.1$--$2.0\times$ faster than the best Base path, the same runtime without residency, retention, or tuned delivery (Figure~\ref{fig:envelope}).

\paragraph{Configuration changes the chosen path.}
At global batch 96, configuration reverses the ranking: EP4 is $1.17\times$ faster before configuration (14.9 versus 17.4\,s), but Local is $1.21\times$ faster afterward (11.4 versus 13.8\,s; Figure~\ref{fig:envelope}). Selecting before configuration would pick EP4 and run $1.21\times$ slower.

\paragraph{Short-test path selection.}
Because decoder layers repeat, tests of one and two repeating units price each path and read its GPU occupancy, and one more test executes the plan's own resident mix, all within 260\,GiB of host memory. Across 16 workloads on four models at 1K and 4K context, the selected path matches full-model runs every time. On six of them, with decisions recorded before full-model execution (Table~\ref{tab:short-test}), predicted EP step time is within 8.3\% (5.6\% on average), and selection follows the crossover: at global batch 192, Local for Gemma4 (predicted $1.16\times$, measured $1.29\times$) and EP4 for Qwen3 ($1.17\times$, $1.08\times$). Searching residency over these measured costs also yields the EP plans for models beyond the calibrated profile (Figure~\ref{fig:generality}a).

\paragraph{Candidate ranking and independent execution.}
At global batch 384 (96 per GPU) and 1K context on RTX 4090, two Local candidates with 16K or 32K GDN/MoE chunks fit a 23\,GiB budget. Predictions recorded before measurement favor 32K (38.66 versus 39.85\,s), as do measured times (40.34 versus 42.13\,s). The selected plan executes in independent processes at 40.38\,s within 21.34\,GiB.

\paragraph{Prediction and planning cost.}
Predictions made before execution are within 5.5\% of measured step time over Local's calibrated 128K--512K range (median 0.1\%) and 0.3\,GiB of host memory; the batch-32 and batch-96 EP4 plans are within 2.3\%. Given each batch's GPU headroom, a complete EP search takes under 9 CPU seconds and ranks the executed plan first or within 0.5\% of its fastest prediction across global batches 16--256; Local searches its full partition menu offline on one core in under an hour.

\subsection{Training Quality}
\label{sec:eval:numerical}
\sys changes where state lives and when operators run, not the training computation: it uses no quantization, pruning, or approximate attention, and checkpointing recomputes the original operators. We train \sys and FSDP2 independently from the same checkpoint and data order. EP4 with 21 resident layers and Local at 1K context (500 steps, 16.4M tokens), and partitioned Local at 32K (128 steps, 16.8M tokens), stay within 0.11\% of FSDP2's held-out NLL at every checkpoint; final NLL differs by at most 0.08\%.

\subsection{Model Coverage and Scale}
\label{sec:eval:models}
\begin{figure}[t]
\centering
\includegraphics[width=\columnwidth]{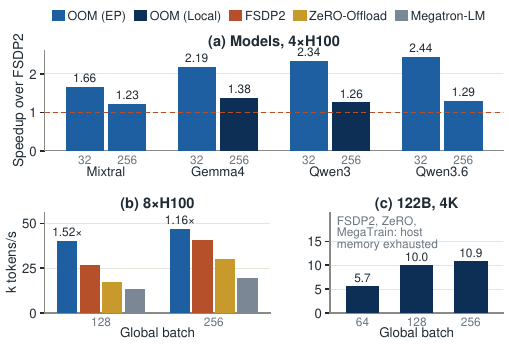}
\caption{Model and node scaling. (a,b) 1K context; (c) Qwen3.5-122B-A10B. (b,c) Thousands of tokens/s; labels in (b): speedup over the fastest baseline.}
\label{fig:generality}
\end{figure}

At 1K context, across four architectures with 8--256 experts, different routing degrees, and full or hybrid attention, \sys delivers $1.66$--$2.44\times$ FSDP2 throughput at batch 32 and $1.23$--$1.38\times$ at batch 256 on four H100s (Figure~\ref{fig:generality}a). All four use EP4 at batch 32; at 256, Gemma4 and Qwen3 use Local, while Mixtral and Qwen3.6 retain EP4 at this lighter token load. Configuration thus adapts to model structure as well as workload size. \sys also trains Qwen3.5-122B-A10B~\cite{qwen2026qwen35config} on four H100s at 4K context, reaching 10.0K tokens/s at global batch 128 and 10.9K at 256 (Figure~\ref{fig:generality}c); at 256, regenerating every other decoder boundary keeps host memory at 1,612\,GiB. Offloaded FSDP2, ZeRO-Offload, and MegaTrain exhaust the node's host memory already at batch 64, before completing a step. With the same FP32 masters and moments, KT-FT exhausts host memory even at global batch 32 and 1K context, with rank 0 alone reaching 1.9\,TiB. Residency also pays on 122B: at 1K context and global batch 16, six resident expert layers make EP4 $1.34\times$ faster and use 162\,GiB less host memory.

On eight H100s at 1K context, \sys with EP8 reaches 40.5K and 47.2K tokens/s at global batch 128 and 256: $1.16$--$1.52\times$ the throughput of offloaded FSDP2, $1.57$--$2.31\times$ that of ZeRO-Offload, and $2.38$--$3.00\times$ that of Megatron-LM (Figure~\ref{fig:generality}b). EP8 serves disjoint expert shards, so per-GPU expert state shrinks with node size.

\section{Conclusion}
\label{sec:conclusion}
When host memory supplies model-state capacity, workload and expert execution determine the value of retaining state and activations, while memory allocation can change which path is faster. Through expert ownership, operator dependencies, and workload-aware configuration, \sys turns GPU and host capacity into throughput. On four H100s, it trains a 122B-parameter model, and a 35B model at 1M-token contexts; on four RTX 4090s, it processes 1,024-sequence batches without accumulation. Across the main batch sweeps, it delivers $1.58\times$ and $4.08\times$ geometric-mean speedups over offloaded FSDP2 on H100 and RTX 4090 with the least host memory among the evaluated systems. Short executions select the faster configured path across 16 workloads on four models. These results make efficient full-parameter MoE training practical on a single node.

\label{p3:main-end}
\makeatletter
\immediate\write\@auxout{\string\newlabel{p4:main-figures}{{\arabic{figure}}{\thepage}{}{}{}}}
\immediate\write\@auxout{\string\newlabel{p4:main-tables}{{\arabic{table}}{\thepage}{}{}{}}}
\immediate\write\@auxout{\string\newlabel{p4:main-equations}{{\arabic{equation}}{\thepage}{}{}{}}}
\makeatother
\clearpage
\label{p3:references-start}
\bibliographystyle{mlsys2025}
\bibliography{references}
\label{p3:references-end}
\label{p3:ai-end}
\clearpage
\appendix
\section{Experimental Setup}
\label{app:setup}
Unless stated otherwise, runs use four GPUs and the Qwen3.6-35B-A3B language backbone, and batch sizes are global. AdamW uses learning rate $10^{-5}$, weight decay 0.01, betas $(0.9,0.999)$, and epsilon $10^{-8}$. $R$ and $K$ count resident expert layers and MoE-retained layers; GAS denotes gradient-accumulation rounds.
\begin{table*}[t]
\centering
\caption{Evaluated systems. All use BF16 compute and FP32 canonical masters and moments.}
\label{tab:system-config}
\begin{tabular}{@{}L{.13\textwidth}L{.28\textwidth}L{.30\textwidth}L{.21\textwidth}@{}}
\toprule
System & Parallelism / canonical state & Activation / transfer & Gradient processing \\
\midrule
\sys & EP4: selected expert state on GPU; Local: host & Workload-specific checkpointing, retention, and partitioning; EP4 full and Local shard transfer & GPU reduction, or BF16 partials with FP32 host accumulation \\
FSDP2 & CPU-sharded state & Layer checkpointing and activation offload & FP32 reduce-scatter \\
ZeRO-Offload & ZeRO-3 with CPU parameter and optimizer offload & Layer checkpointing and activation offload & BF16 communication \\
Megatron-LM & TP2/SP, EP4/ETP1, DP2, PP1; CPU optimizer & Full recomputation & FP32 main gradients \\
MegaTrain & Layer streaming with CPU offload & Its streaming path with FA3 & Host optimizer \\
KT-FT & Routed experts on CPU (AMX); others through FSDP2 & Recompute; official full-parameter path & BF16 expert gradients, FP32 update \\
\bottomrule
\end{tabular}
\end{table*}

\begin{table}[t]
\centering
\small
\setlength{\tabcolsep}{4pt}
\caption{Routed-expert geometry of the evaluated language backbones~\cite{jiang2024mixtral,qwen2025qwen3config,qwen2026qwen36config,qwen2026qwen35config,gemmateam2026gemma4}.}
\label{tab:model-geometry}
\begin{tabular}{@{}lrrr@{}}
\toprule
Model & Experts $E$ & Active $k$ & $E/k$ \\
\midrule
Mixtral-8$\times$7B & 8 & 2 & 4 \\
Gemma4-26B-A4B & 128 & 8 & 16 \\
Qwen3-30B-A3B & 128 & 8 & 16 \\
Qwen3.6-35B-A3B & 256 & 8 & 32 \\
Qwen3.5-122B-A10B & 256 & 8 & 32 \\
\bottomrule
\end{tabular}
\end{table}

\paragraph{Platforms and CPU placement.}
Table~\ref{tab:system-config} summarizes the evaluated systems, and Table~\ref{tab:model-geometry} the routed-expert geometry of the evaluated models. The H100 host has eight NVLink-connected 80\,GB GPUs and two 48-core Xeon Platinum 8468 sockets with 1,008\,GiB of memory each. The RTX 4090 host has four 24\,GB PCIe GPUs and two 32-core Xeon Gold 6338N sockets with 503\,GiB each. Each comparison runs all systems on the same GPUs and host; on H100, every system runs its CPU work on the socket local to its GPUs.

\paragraph{Measurement.}
Each configuration runs as independent processes. A measured step covers forward, replay, backward, communication, optimizer updates, and weight publication, and ends after pending transfers finish, CUDA synchronizes, and all ranks reach a barrier. We take the slowest rank at each step and the median within each process. Main sweeps report the mean of two process medians, each over two warmup and five measured steps; mechanism studies (Figures~\ref{fig:c2-ablation}a and~\ref{fig:c2-policies}) use one process with one warmup and three measured steps. Host memory is the peak PSS of the training process tree; for 512K/1M and the mechanism studies, it is the runtime's rank-level PSS sum. GPU memory is the largest allocated peak across ranks.

\section{Execution Settings}
\label{app:execution}
\paragraph{Expert-owned state execution.}
EP4 uses two managed GPU buffers and two CPU gradient slots each for managed parameters and expert shards, with a dedicated stream for expert-gradient D2H. Multithreaded CPU AdamW updates host state; GPU AdamW updates resident expert state and refreshes its BF16 weights. Managed parameters and expert shards share one update executor per rank. MoE-retained layers keep their backward tensors on GPU and replay only attention; other layers checkpoint the full decoder with host-stored inputs.

\paragraph{Split--Keep--Replay storage.}
Retained attention outputs $O$, log-sum-exp statistics (LSE), and decoder inputs live in pinned host pools whose NUMA placement is part of the configuration. A retained object is written once during the original forward, read by replay and backward, and released after its last reader completes; reuse waits for that completion. GDN recurrent states connect the chunks of one layer and remain on GPU through that layer's backward. Regeneration reuses the retained input's prefetched copy and keeps the last traversed decoder's parameters loaded for its backward.

\section{Executed Configurations}
\label{app:configs}
\label{app:configuration}
\paragraph{H100 batch sweep.}
Table~\ref{tab:main-configurations} lists the executed H100 4K configurations. EP4 uses full parameter transfer and two CPU gradient slots; Local uses full-layer checkpointing with host-stored inputs, shard transfer with 256\,MiB chunks, and one CPU gradient slot. All use GAS 1; batch 32 runs the planner's proposal, with 23 interleaved resident layers and 64K-token routing chunks.

\begin{table}[t]
\centering
\small
\caption{Executed configurations in the H100 4K sweep.}
\label{tab:main-configurations}
\begin{tabular}{@{}rllr@{}}
\toprule
Batch & Path & State / activation & Step (s) \\
\midrule
16 & EP4 & R27/K0 & 3.776 \\
32 & EP4 & R23/K0 & 5.020 \\
64 & EP4 & R3/K4 & 9.302 \\
128 & Local & Host state, checkpoint offload & 12.177 \\
256 & Local & Host state, checkpoint offload & 22.305 \\
\bottomrule
\end{tabular}
\end{table}

\paragraph{RTX 4090 batch sweep.}
At global batches 64--1024 (1K context), \sys uses per-GPU batches 16--256 with GAS 1; FSDP2 and ZeRO-Offload use microbatches 16/32/32/32/32 with GAS 1/1/2/4/8; MegaTrain, whose multi-GPU path does not accumulate gradients, exhausts GPU memory from 32 sequences per GPU. Up to 64 sequences per GPU, \sys uses the default operators. At 128 per GPU, GDN and MoE run in 32K- and 16K-token partitions; at 256 per GPU, the Local planner's plan uses 16K-token GDN, 32K-token MoE, and 64K-token attention batch partitions.

\paragraph{Long sequences.}
With one sequence per GPU, 64K and 128K use the default operators. At 256K, Split--Keep--Replay uses query group 2, 128K-token GDN and 256K-token MoE partitions, and retains every boundary and $O$/LSE; it takes 62.80\,s versus 66.86\,s for the default path, with allocated peak 39.4 versus 53.8\,GiB. The 512K and 1M configurations use YaRN scaling factors two and four, 1+1 and 3+1 regeneration, and take 205.66 and 712.64\,s per step. At 512K, FSDP2 exhausts GPU memory during GDN checkpoint replay, MegaTrain in GDN forward with checkpoint intervals of four and six, and ZeRO-Offload in its first step; all three do so again at 1M.

\paragraph{Megatron-LM at 4K context.}
\label{app:megatron-s4k}
Megatron-LM uses TP2 with sequence parallelism, EP4/ETP1, PP1, and dense DP2, with full recomputation, vocabulary-parallel cross-entropy, and a CPU optimizer holding FP32 masters, gradients, and moments. Among the tested microbatch/GAS pairs, 16/1 exhausts memory in the first output projection and 4/4 in cross-entropy. The completing 2/8 setting takes 23.93\,s per step (5.48K tokens/s) with 68.0\,GiB allocated; our EP4 plan's 5.020\,s is $4.77\times$ faster.

\paragraph{KT-FT.}
KT-FT runs its official Accelerate and LlamaFactory full-parameter path with native AMX BF16 CPU experts on one 48-core socket (44 expert workers, the fastest of 40, 44, and 48; 24 Torch/BLAS threads; recompute). Our changes add compatibility fixes, packed inputs, timing, and FP32 master weights and Adam moments. Step times at global batch 16, 32, 64, and 128 are 99.49, 168.38, 326.55, and 726.60\,s; batch 256 exhausts GPU memory at GAS 1 and takes 1,396.19\,s at GAS 2. With one sequence per GPU, 64K and 128K take 314.29 and 734.61\,s; 256K exhausts GPU memory with both the default and expandable allocators. Sampled host peaks are 619, 656, 732, 876, and 876\,GiB at batches 16--256. At batch 32, the CPU expert forward, recompute, and backward paths take 70.2\% of rank 0's exposed wall time and the FP32 update 4.2\%. For Qwen3.5-122B-A10B, FP32 masters and moments add 864\,GiB over KT-FT's native BF16 optimizer state; at global batch 32 and 1K context, the run exhausts the node's host memory before its first update, with 1,946\,GiB of anonymous memory on rank 0.

\paragraph{Base and configured paths.}
Figure~\ref{fig:envelope} measures each layout before and after configuration. Base uses full parameter transfer, 64\,MiB H2D/D2H chunks, full checkpoint offload, and one CPU gradient slot; EP4 Base has no resident layers and 32K-token routing chunks. Configured EP4 sets resident layers, MoE retention, and 32K--256K-token routing chunks for each batch, with two CPU gradient slots and the dedicated gradient stream; configured Local uses shard transfer with 256--512\,MiB chunks. At batch 96, configured EP4 uses R1/K1.

\paragraph{Path comparison in Figure~\ref{fig:intro-pipeline}c.}
EP4/Local step times are 5.020/10.035, 9.302/10.245, and 18.714/12.177\,s at global batch 32, 64, and 128. EP4 uses R23/K0, R3/K4, and R1/K1 with full transfer and two gradient slots; Local uses shard transfer with one slot.

\section{Mechanism Measurements}
\label{app:mechanisms}
\paragraph{Residency and retention.}
The batch-32/4K residency pair in Section~\ref{sec:eval:mechanisms} runs the planned EP4 configuration with no resident layers (8.48\,s) and with the planned 23 (5.02\,s), raising GPU allocation from 11.4 to 64.8\,GiB. The gain persists at 1K context, where 27 resident layers reduce step time from 8.11 to 3.54\,s ($2.29\times$). Interleaved sets keep decoders 0 and 1 resident, which finish backward last and are needed first by the next forward, and space the rest evenly: the 12-layer pair (64K routing chunk) compares decoders 0--11 with decoders 0, 1, 3, 7, 11, 15, 19, 22, 26, 30, 34, and 38. Table~\ref{tab:residency-retention} gives the residency--retention pairs in Section~\ref{sec:eval:mechanisms} at global batch 32 and 4K context, two processes each. R21 adds four resident layers (decoders 33, 35, 37, and 39) to R17; K1 retains decoder 0's MoE activations while its attention still replays.

\begin{table}[t]
\centering
\small
\caption{EP4 residency and retention at batch 32, 4K: step time (s) and allocated peak (GiB).}
\label{tab:residency-retention}
\begin{tabular}{@{}lrrrr@{}}
\toprule
 & \multicolumn{2}{c}{K0} & \multicolumn{2}{c}{K1} \\
\cmidrule(lr){2-3}\cmidrule(l){4-5}
Resident & Time & Peak & Time & Peak \\
\midrule
R17 & 6.048 & 51.3 & 5.733 & 55.0 \\
R21 & 5.376 & 60.3 & 5.331 & 64.1 \\
\bottomrule
\end{tabular}
\end{table}

\paragraph{Cumulative 512K ablation.}
\label{app:c2-ablation}
All arms use Local execution with shard transfer in 256\,MiB chunks, GPU gradient reduction, full checkpoint offload, one CPU gradient slot, and YaRN$\times2$. The default arm uses the unpartitioned operators and exhausts GPU memory in the first MoE backward. +Split groups eight query heads, partitions GDN and MoE into 128K- and 256K-token chunks, releases GDN intermediates at last use, skips unused terminal MoE outputs during replay, and defers backward input prefetch until attention workspace is released. +Keep retains forward $O$/LSE on the host. +Replay keeps decoder inputs $0,2,\ldots,38$ and regenerates the rest, holding at most one regenerated boundary. Table~\ref{tab:c2-ablation} gives the values in Figure~\ref{fig:c2-ablation}a.

\begin{table}[t]
\centering
\small
\caption{Cumulative Split--Keep--Replay arms at 512K: step time (s), GPU allocated peak and host memory (GiB).}
\label{tab:c2-ablation}
\begin{tabular}{@{}lrrr@{}}
\toprule
Arm & Time & GPU & Host \\
\midrule
Default & Out of GPU memory & --- & --- \\
+ Split & 260.02 & 43.4 & 731.3 \\
+ Keep & 200.46 & 43.4 & 892.5 \\
+ Replay & 206.00 & 45.4 & 732.5 \\
\bottomrule
\end{tabular}
\end{table}

\paragraph{Policies by length.}
Table~\ref{tab:c2-policies} gives the values in Figure~\ref{fig:c2-policies}. A retains every decoder boundary and recomputes attention; B also retains $O$/LSE; 1+1 and 3+1 retain $O$/LSE and keep one boundary in every two and four layers. Policies share the 512K operator partitions, clamped to the sequence length; at 512K, A and B are the +Split and +Keep arms above. The length-dependent benefit of retention and the host savings of regeneration make the best policy workload-specific. Under a 750\,GiB budget at 512K, the Local planner stores boundaries 0--21 on the host, regenerates boundary 22, and holds boundaries 23--39 on GPU; this plan takes 202.06\,s (predicted 201.66\,s) at 63.8\,GiB of GPU allocation.

\begin{table}[t]
\centering
\small
\setlength{\tabcolsep}{4pt}
\caption{Policies by length: step time (s) / host memory (GiB).}
\label{tab:c2-policies}
\begin{tabular}{@{}lrrrr@{}}
\toprule
Length & A & B & 1+1 & 3+1 \\
\midrule
32K & 10.69/431 & 11.70/441 & 11.75/431 & 14.62/426 \\
64K & 12.32/451 & 11.62/471 & 12.94/451 & 15.17/441 \\
128K & 26.48/491 & 23.27/531 & 26.46/491 & 26.29/471 \\
256K & 77.83/571 & 62.55/651 & 66.85/569 & 66.88/531 \\
512K & 260.0/731 & 200.5/893 & 205.5/732 & 208.2/652 \\
\bottomrule
\end{tabular}
\end{table}

\paragraph{Sliding-window hybrid.}
\label{app:gemma4-costs}
Gemma4-26B-A4B interleaves five sliding-window layers (1K window, head dimension 256) with each global-attention layer (head dimension 512, two shared key/value heads). Table~\ref{tab:gemma4-costs} gives per-layer forward times for one sequence on one H100 and the host payload of each 5+1 block. Global attention uses PyTorch's memory-efficient kernel, since FlashAttention does not support head dimension 512.

\begin{table}[t]
\centering
\small
\setlength{\tabcolsep}{3.2pt}
\caption{Gemma4-26B-A4B per-layer forward time (s) on one H100 and host payload per 5+1 block (GiB).}
\label{tab:gemma4-costs}
\begin{tabular}{@{}lrrrrrr@{}}
\toprule
Length & Global & Reuse $O$ & Sliding & MoE & $O$/LSE & Freed \\
\midrule
32K & 0.160 & 0.004 & 0.005 & 0.015 & 0.50 & 0.86 \\
64K & 0.821 & 0.008 & 0.013 & 0.030 & 1.00 & 1.72 \\
128K & 3.904 & 0.021 & 0.021 & 0.071 & 2.01 & 3.44 \\
256K & 17.164 & 0.039 & 0.042 & 0.147 & 4.02 & 6.88 \\
\bottomrule
\end{tabular}
\end{table}

\paragraph{Largest completed workloads.}
For Figure~\ref{fig:c2-ablation}b, the H100 context endpoint is the 1M configuration above, and the H100 batch endpoint runs Local at 256 sequences of 4K per GPU with 64K-token GDN, MoE, and attention batch partitions, at 66.3\,GiB peak allocation. The RTX 4090 endpoints are the 256-per-GPU plan above and one 512K sequence per GPU with grouped attention and GDN/MoE partitions, at 21.95\,GiB peak allocation. The default path completes 256K and 64 sequences per GPU on H100, and 64K and 64 per GPU on RTX 4090; each exhausts GPU memory at the next doubling. A single H100 with the same mechanisms and YaRN factor eight completes a 2M-token step in 2,606\,s, with 67.0\,GiB of GPU allocation and 915\,GiB of host memory.

\section{Configure-then-Compare Details}
\label{app:planning}
\paragraph{Cost and memory accounting.}
EP calibration scales GPU work with token count, treats parameter work as approximately token-independent, and adds a per-chunk scheduling charge. The Local time model sums operator time, boundary regeneration, exposed transfer chains, and a fixed overhead. GPU feasibility sums the allocations live in each execution phase and takes the largest phase total; views and their backing storage are counted once. Host estimates include model state, pinned pools, checkpoints, retained results, and staging, charged to the NUMA node of each allocation.

\paragraph{Selection rules and plan output.}
Local prices each candidate in under a millisecond and searches its full partition menu offline on one core in under an hour. Among candidates within 0.5\% of the minimum predicted time, the EP search prefers the one with the most memory headroom. A saved plan records the layout, resident and retained layer indices, operator chunks, parameter-transfer mode and chunk sizes, buffer settings, the boundary keep set, host placement, and resource estimates; the runtime executes it directly.

\paragraph{Prediction accuracy.}
Table~\ref{tab:prediction} lists plans whose predictions were recorded before execution; the long-sequence rows cover 128K--512K, the range on which the Local time model is calibrated. With operator partitions and $O$/LSE retention fixed at 512K, the model predicts 205.70 and 207.89\,s for 1+1 and 3+1 regeneration, respectively, matching their measured order at 205.49 and 208.18\,s. The five EP4 budget policies at batch 32/4K (64 and 72\,GiB budgets) are predicted within 3.3--6.6\% of measured time and 0.24\,GiB of allocated memory.

\begin{table}[t]
\centering
\small
\setlength{\tabcolsep}{4pt}
\caption{Predicted versus measured step time (s), with predictions recorded before execution. Error is predicted/measured $-1$.}
\label{tab:prediction}
\begin{tabular}{@{}llrrr@{}}
\toprule
Workload & Plan & Pred. & Meas. & Error \\
\midrule
H100, B32, 4K & EP4 R23 & 5.135 & 5.020 & $+2.3\%$ \\
H100, B96, 4K & EP4 R1/K1 & 13.73 & 13.78 & $-0.4\%$ \\
4090, B384, 1K & Local, 32K chunks & 38.66 & 40.38 & $-4.3\%$ \\
H100, 128K & B & 23.21 & 23.27 & $-0.2\%$ \\
H100, 128K & 3+1 & 24.85 & 26.29 & $-5.5\%$ \\
H100, 256K & B & 62.73 & 62.76 & $-0.1\%$ \\
H100, 512K & 1+1 & 205.70 & 205.49 & $+0.1\%$ \\
H100, 512K & 3+1 & 207.89 & 208.18 & $-0.1\%$ \\
\bottomrule
\end{tabular}
\end{table}

\paragraph{Short-test path selection.}
Table~\ref{tab:short-test} lists the 16 workloads in Section~\ref{sec:eval:selection}, on four H100s. Each test runs one warmup and five measured steps on whole repeating units: four and eight layers, or six and twelve for Gemma4's 5+1 block. Decisions for the workloads marked $\dagger$ were recorded before their full-model runs.

\begin{table}[t]
\centering
\small
\setlength{\tabcolsep}{4pt}
\caption{Predicted/measured step time (s) of both configured paths; the selected path is in bold.}
\label{tab:short-test}
\begin{tabular}{@{}lllccc@{}}
\toprule
Model & $S$ & Batch & Plan & EP4 & Local \\
\midrule
Qwen3.6 & 4K & 48 & R15/K1 & \textbf{7.08/7.29} & 9.27/10.20 \\
Qwen3.6 & 4K & 192 & R1 & 25.91/26.72 & \textbf{17.09/17.51} \\
Qwen3.6 & 1K & 32 & R27 & \textbf{4.03/3.54} & 10.20/10.35 \\
Qwen3.6$^\dagger$ & 1K & 128 & R23 & \textbf{5.54/5.12} & 10.27/10.26 \\
Qwen3.6 & 1K & 256 & R19 & \textbf{8.54/9.29} & 9.55/10.08 \\
Mixtral & 1K & 32 & R15 & \textbf{6.26/6.31} & 11.62/12.21 \\
Mixtral$^\dagger$ & 1K & 128 & R11 & \textbf{9.23/10.06} & 14.30/14.68 \\
Mixtral & 1K & 256 & R7 & \textbf{16.85/18.06} & 21.78/21.67 \\
Gemma4 & 1K & 32 & R27 & \textbf{2.78/2.89} & 7.87/7.79 \\
Gemma4$^\dagger$ & 1K & 128 & R18 & \textbf{6.98/7.07} & 8.24/8.34 \\
Gemma4$^\dagger$ & 1K & 192 & R8 & 9.91/10.17 & \textbf{8.51/7.90} \\
Gemma4 & 1K & 256 & R1 & 13.78/13.32 & \textbf{9.79/9.45} \\
Qwen3 & 1K & 32 & R36 & \textbf{3.19/2.95} & 7.77/8.12 \\
Qwen3$^\dagger$ & 1K & 128 & R30 & \textbf{5.32/5.63} & 8.05/8.87 \\
Qwen3$^\dagger$ & 1K & 192 & R26 & \textbf{7.11/7.72} & 8.30/8.34 \\
Qwen3 & 1K & 256 & R18 & 10.04/10.55 & \textbf{8.93/8.90} \\
\bottomrule
\end{tabular}
\end{table}

\section{Training Quality}
\label{app:quality-protocols}
\begin{figure}[t]
\centering
\includegraphics[width=\columnwidth]{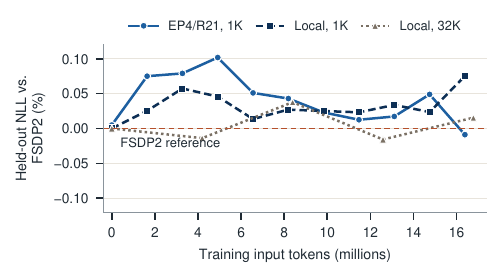}
\caption{Held-out NLL gap to independent FSDP2 runs from the same checkpoint and data order for three configurations of our system.}
\label{fig:release-quality}
\end{figure}

\sys and FSDP2 train independently from the same checkpoint and data order; held-out NLL is measured on token- and block-disjoint data. Local and EP4/R21 at eight 1K sequences per GPU run 500 steps (16.384M input tokens) with evaluation at step 0 and every 50 steps; Local with partitioned operators at one 32K sequence per GPU (query group 2, GDN 8K, MoE 4K) runs 128 steps (16.777M tokens) with evaluation every 32 steps. Evaluation runs between steps and restores the training state before training resumes. Held-out NLL decreases at every checkpoint for both systems, and the relative gap stays within 0.11\%. With the same seed and inputs, policies B, 3+1, and 1+1 (Table~\ref{tab:c2-policies}) match policy A's first-step loss to four decimals at 32K and 128K and stay within 0.0006 of it over the next three steps. Final \sys/FSDP2 NLL is 1.376510/1.375475 for Local at 1K, 1.375354/1.375475 for EP4, and 1.159174/1.158994 at 32K (Figure~\ref{fig:release-quality}).

\section{Models and Eight GPUs}
\label{app:models}
\paragraph{Additional architectures.}
Mixtral-8x7B (32 layers, eight experts, top-2, 4K sliding window), Gemma4-26B-A4B (30 layers, 128 experts, top-8, with a dense MLP beside the experts), and Qwen3-30B-A3B (48 layers, 128 experts, top-8) run on four H100s at 1K context. At batch 32, \sys uses EP4 with R15 (Mixtral), R27 (Gemma4), and R36 (Qwen3-30B) resident layers; step times versus FSDP2 are 6.308/10.503, 2.887/6.318, and 2.954/6.918\,s. At batch 256, \sys uses EP4 for Mixtral and Local for Gemma4 and Qwen3-30B; step times versus FSDP2 are 18.057/22.129, 9.454/13.076, and 8.897/11.247\,s. FSDP2 uses the same expert kernel as \sys, or its own faster Liger kernel.

\paragraph{Qwen3.5-122B-A10B.}
The 48-layer model (256 top-8 experts, hidden size 3072) runs on four H100s. Local execution at 4K context, without resident layers, takes 46.0, 52.3, and 96.3\,s per step at global batch 64, 128, and 256 (5.7K, 10.0K, and 10.9K tokens/s), with 44.9--65.4\,GiB of peak GPU allocation and 1,516--1,612\,GiB of host memory. Batch 256 uses 64K-token GDN, MoE, and attention batch partitions and regenerates every other decoder boundary, since the pinned host memory for keeping every boundary exceeds what our test machine can allocate. With EP4 at 1K context and global batch 16, six resident expert layers reduce step time from 55.7 to 41.7\,s ($1.34\times$) and host memory from 1,420 to 1,258\,GiB. CPU-offloaded FSDP2 (batch 64, microbatch 4, GAS 4) uses over 1.8\,TiB of host memory and exhausts the node before completing its first step; ZeRO-Offload with the same settings exhausts it during setup at 1.82\,TiB, and MegaTrain at batch 64 exhausts it during its first step. FSDP2 and ZeRO-Offload keep FP32 gradients, master weights, and Adam moments in host memory, at least 16 bytes per parameter (about 1.8\,TiB for this model).

\paragraph{Eight GPUs.}
The eight-H100 comparison uses 1K context and global batches 128 and 256, with each rank on its local NUMA node. \sys uses EP8/R21 with GAS 1. FSDP2 uses the CPU-offloaded recipe of the four-GPU model comparison and ZeRO-Offload its CPU offload, both with microbatches 16 and 32; Megatron-LM keeps its CPU optimizer and uses microbatch 32 with GAS 2 and 4, since microbatch 64 exhausts GPU memory. All use FA2 attention. Step times at batch 128 and 256 are 3.238/5.548\,s for \sys, 4.932/6.445\,s for FSDP2, 7.482/8.732\,s for ZeRO-Offload, and 9.713/13.232\,s for Megatron-LM.

\label{p3:appendix-end}
\end{document}